\documentclass[twocolumn]{aa}
\begin{document}

\title{The Swift satellite and redshifts of long gamma-ray bursts}

\titlerunning{The Swift satellite and redshifts ...}

\author{Z. Bagoly \inst{1}
\and
A. M\'esz\'aros \inst{2}
\and
L.G. Bal\'azs \inst{3}
\and
I. Horv\'ath \inst{4}
\and
S. Klose \inst{5}
\and
S. Larsson \inst{6}
\and
P. M\'esz\'aros \inst{7}
\and
F. Ryde \inst{6}
\and
G. Tusn\'ady \inst{8}
 }

\offprints{A. Bagoly}

\institute{Laboratory for Information Technology, E\" otv\" os
          University, H-1117 Budapest, P\'azm\'any P. s. 1/A,
          Hungary\\
          \email{zsolt.bagoly@elte.hu}
\and
Astronomical Institute of the Charles University,
              V Hole\v{s}ovi\v{c}k\'ach 2, CZ 180 00 Prague 8,
          Czech Republic\\
              \email{meszaros@mbox.cesnet.cz}
\and
              Konkoly Observatory, H-1525 Budapest, POB 67, Hungary\\
    \email{balazs@konkoly.hu}
\and
          Department of Physics, Bolyai Military University, H-1456
                           Budapest, POB 12, Hungary\\
           \email{horvath.istvan@zmne.hu}
\and
     Tautenburg Observatory, D-07778 Tautenburg, Sternwarte 5,
         Germany \\
          \email{klose@tls-tautenburg.de}
\and
       Stockholm Observatory, AlbaNova, SE-106 91 Stockholm, Sweden\\
         \email{stefan@astro.su.se,felix@stro.su.se}
\and
       Dept. of Astronomy \& Astrophysics, Pennsylvania State
                 University, 525 Davey Lab. University Park, PA 16802, USA\\
    \email{nnp@astro.psu.edu}
\and
    R\'enyi Institute of Mathematics, Hungarian Academy of Sciences,
     H-1364 Budapest, POB 127, Hungary\\
         \email{tusnady@renyi.hu}
              }

   \date{Received 7 October  2005; accepted  ..........}

\abstract{Until 6 October 2005 sixteen redshifts have been measured of
long gamma-ray bursts discovered by the Swift satellite.  Further $45$
redshifts have been measured of the long gamma-ray bursts discovered by
other satellites. Here we perform five statistical tests comparing the
redshift distributions of these two samples assuming - as the null
hypothesis - identical distribution for the two samples. Three tests
(Student's $t$-test, Mann-Whitney test, Kolmogorov-Smirnov test) reject
the null hypothesis on the significance levels between $97.19$ and
$98.55$\%. Two different comparisons of the medians show extreme
$(99.78-99.99994)$\% significance levels of rejection. This means that
the redshifts of the Swift sample and the redshifts of the non-Swift
sample are distributed differently - in the Swift sample the redshifts
are on average larger. This statistical result suggests that the long
GRBs should on average be at the higher redshifts of the Swift sample.

}
\maketitle

\section{Introduction}

Recently, the Swift satellite (\cite{swift}) detected the gamma-ray
burst GRB050904, for which the redshift is $z=6.29$, directly measured
from the afterglow data (\cite{gcn3914,gcn3919,gcn3924,gcn3937}).  In
addition, it is remarkable  
that the GRBs detected by the Swift satellite 
seem to have systematically bigger redshifts on average
than the redshifts of GRBs detected by other satellites
(\cite{gre05,fri05}).  Hence, the question emerges immediately: does the
redshift distribution of GRBs detected by Swift significantly differ
from the other GRBs with known redshifts?

The search for the answer to this question is the first aim of this 
paper: using statistical analyses we show that this is really the case.
We briefly discuss the importance, meaning and the consequence of this
statistical result on the redshift distribution of long GRBs.  The
detailed discussion of the astrophysical reasons 
for the obtained results is the topic of a forthcoming paper.

\section{The samples}

Greiner (2005) lists the observations concerning the afterglows of GRBs,
and among them he also selects and lists the confirmed redshifts (see
the reference therein and also \cite{fri05}).  For our statistical
studies this survey and selection was used.  There are 16 bursts with
measured redshifts, which were detected first by the Swift satellite,
during the period 1 January - 6 October 2005. These 16 redshifts define
here the "Swift" sample, where the smallest redshift is $z=0.26$
(GRB050714, \cite{gcn3679}); the largest redshift is $z=6.29$
(GRB050904, \cite{gcn3937}); the mean redshift is
$\overline{z}_{\mbox{\em Swift}}= 2.42$; the variance is
$\sigma^2_{\mbox{\em Swift}}= 2.60$; and the median is $z_{\mbox{\em
Swift,med}}= 2.405$.

There are a further 45 redshifts in the Greiner's table from the period 28
February 1997 - 6 October 2005.  In the cases GRB011030X and GRB980329
there are upper limits only, and these cases were not included into the
sample; in three cases (GRB020305; GRB991216; GRB980326) the redshifts
were only estimated, but here we considered them; for GRB000214 the
estimated redshift is between 0.37 and 0.47, and here $z=0.42$ was used
(\cite{antonelli2000}). In this way we obtain the "non-Swift" sample
containing 45 redshifts.  In this sample the smallest redshift is
$z=0.0085$ (GRB980425, \cite{galama98}); the largest redshift is
$z=4.50$ (GRB000131, \cite{andersen2000}); the mean redshift is
$\overline{z}_{\mbox{\em non-Swift}}= 1.31$; the variance is
$\sigma^2_{\mbox{\em non-Swift}}= 1.11$; and the median is 
$z_{\mbox{\em non-Swift,med}}= 1.02$.

It should be noted that both samples probably only constitute long GRBs
($ > 2s$), which should be different from short and intermediate bursts
(\cite{ho98,mee01,ho02,bal03,hak03,bor04}). In other words, this paper
deals exclusively with long GRBs.

\section{The statistical tests and their results}

The two samples were compared using the parametric Student's $t$-test and
non-parametric rank based statistical tests: the Mann-Whitney test,
Kolmogorov-Smirnov test and two median tests. The rank based tests has
the clear advantage of being unaffected by any monotonic transformation
in the $z$ values.

The most common test for the comparison of the average redshifts is the
Student's $t$-test. The details of this test together with the
calculation of the significance level can be found in Chap. 14.2 of
Press et al. (1992). Note here that, since the variances are different,
one has to use the modification of the test for unequal variances.  The
relevant formulas are given by Eq.(14.2.3) and Eq.(14.2.4).  One obtains
$t=2.57$  for the degree-of-freedom {\em dof}$=19.75$.  The Student's
$t$-test's significance of rejecting the null hypothesis (i.e. 
the two samples have identical mean values) is $98.15\%$.

The Mann-Whitney test uses, instead of the values of the redshifts,
their ranks.  Its meaning with the relevant formulas is explained, for
instance, by Lowry (1999).  Not going into the details we remark only
that the key idea of the test is the following: one considers first the
two samples together, and in this common sample, containing in our case
61 objects, one sorts the redshifts into a monotonic increasing
sequence. Having this, not the values of redshifts, but only the ranks
will be used for both in the non-Swift and in the Swift sample.  This
means that the null hypothesis is the assumption that in both samples
the mean ranks are identical, and the difference comes from chance.

In our case the Mann-Whitney test gives a $98.55\%$ significance level, i.e.
it rejects the null hypothesis almost at the same significance level as
that of the Student's $t$-test.

The Kolmogorov-Smirnov test compares the cumulative distributions of the
redshifts in the two samples. The details of the test are described in
Chap.  14.3 of Press et al. (1992). It gives the significance $97.19$\%,
which means that this test rejects the null hypothesis at this level. 

Not only the mean redshifts in the two samples, but also the medians
could be different. To verify this expectation we provided the following
two tests.

We considered the two samples with all together $61$ redshifts, sorted
in ascending series.  There are $6$ non-Swift and $8$ Swift GRBs with
redshifts larger that $z_{\mbox{\em Swift,med}}=2.405$; there are then
$39$ non-Swift and $8$ Swift GRBs with redshifts smaller than
$z_{Swift,med}$. In other words, the median of Swift sample separates
the non-Swift in a ratio $6:39$, instead of the expected $1:1$.

Assuming a binomial distribution with $N = 45$ and $p=0.5$ the expected
most probable value is $N p= 22.5$ (e.g. \cite{me97,bal98}) and the
expected dispersion is $\sigma^2_{\mbox{\em theor.}} = Np(1-p) = 11.25$;
i.e.  $\sigma_{\mbox{\em theor.}} = 3.4$. The difference between the
most expected and the real value is $39 - 22.5 = 16.5 = 4.9
\sigma_{\mbox{\em theor.}}$. Because both $Np$ and $\sigma_{\mbox{\em
theor.}}$ are much greater than one, the Gaussian approximation of the
binomial distribution is allowed. This means that the null hypothesis
(i.e. the two medians are the same) is rejected on the $4.9$ sigma
level, corresponding to the significance $99.99994$\%.

A modification of this test compares the medians of the Swift and
non-Swift samples as follows. Be chosen 16 objects randomly from the
sample of $N=61$ events. In this subsample the mean between the $8^{th}$
and $9^{th}$ objects gives the median. Let this $8^{th}$  ($9^{th}$)
object from the subsample be in the $q^{th}$ ($s^{th}$) position in the
ordered sample of $N$ events ($q <s$). Obviously, it can be $8 \leq q
\leq 53$, and  $9 \leq s \leq 54$.  Then the corresponding probability
for the given $q$ and $s$ is simply the number of the good
configurations divided by the number of all configurations: $P(q,s) =
(\begin{array}{cc} q-1 \\7  \end{array} ) (\begin{array}{cc}{N-s} \\
7\end{array} )/ (\begin{array}{cc}{N}\\16 \end{array} )$, where for
non-negative integer $x$ and $y\geq x$ we have
$(\begin{array}{cc}{y}\\{x}\end{array}) = y!/(x!(y-x)!)$. For the median
of the Swift sample the corresponding significance level is $\sum_{(q+s)
< 95} P(q,s)=99.78$\%. (The sum must be taken for all combinations with
$(q+s)/2 < 47.5$, because for these combinations the medians of the
chosen subsamples are smaller than the Swift's median.) The null
hypothesis is again {\it rejected} on this level.  We verified this high
significance level also by 100000 Monte-Carlo simulations, and confirmed the result above.

\section{Discussion and conclusions}

All five statistical tests  reject the null hypothesis that the redshift
distributions for the Swift and non-Swift samples are identical on the
significance higher than $95$\%.  The redshifts of the Swift sample are
on average larger than that of the non-Swift sample.

Three questions emerge immediately from this result: I. Is this
statistical result an important new result indeed? II. What is the
reason for this behavior? III. And what is the impact of this result on
the redshift distribution of the long GRBs?  

First of all, it may seem that the result is in essence not unexpected
and is "reasonable". This point of view may follow from the fact that
the Swift satellite is surely more sensitive, and it allows for a faster
detection of fainter bursts than that of the other satellites. Also the
number of detected GRBs during a given time interval is larger. Hence,
because "Swift detects more and fainter GRBs, it is reasonable that it
also detects higher redshifts on average" - one may think. 

Concerning this point of view two important remarks are needed. 

First, it is a triviality that either an exact confirmation or an
exact rejection of a theoretical expectation - done purely from the
observational data by strict statistical methods - is useful. This was
done here. In fact, it is not so sure immediately without a detailed
statistical testing that the redshifts of the Swift sample are in
average bigger than that of the non-Swift sample.  
For example, if one takes {\it only} six typical redshift values
(minimal, mean and maximal redshifts from both samples) and one does
{\it not} take also the variances, then one simply cannot obtain from
these six redshifts {\it alone} the conclusion that the two
distributions have different redshifts in average. The six typical
redshifts {\it alone} are simply not enough for such a claim.  
It is not even clear what is the ``most natural'' parameter of the
distance: $z$, $1+z$, $\log(1+z)$ or even the luminosity distance?
Either could be a ``good'' parameter: however they will show totally
different distribution (mean, variance, skewness, etc.), and it is hard
to use parametric tests only.  The rank based tests are clearly
preferred as they have the clear advantage of being free from such
problems.  The relevance of our result for the redshift distribution of
long GRBs is straightforward: it excludes the possibility that the Swift
sample and the non-Swift sample of GRBs originate, on average, in the
same redshift range. All this answers the question I.

Second, one should point out that, strictly from the observational point of
view, it is simply not yet certain that fainter, long bursts are on
average at larger redshifts.  The redshifts of long GRBs are still
poorly measured.  Beyond a few dozens of directly measured redshifts -
the ones studied in this paper - only indirect estimations are known for
the redshifts of long GRBs. The estimates from the BATSE data using
the gamma range alone (\cite{MM96,  ho96, reme07,  lare00,  schmidt2001,
scha01, lr02, no02, bag03, at03, lin04}) suggest that large redshifts
(up to $z=20$ (\cite{MM96})) for fainter GRBs may well be expected
indeed. Also the analyses from the afterglow data suggest (\cite{be05})
that  the fainter, long GRBs really should on average be at higher
redshifts.  The very high GRBs' redshifts are discussed in detail by
Lamb \& Reichart (2000); the cosmological aspects of GRBs - fainter ones
expected to be at higher redshifts - are studied, e.g., by Friedman \&
Bloom (2005). Hence, the theoretical expectation ``fainter = larger
redshifts on average'' is reasonable indeed. But, strictly, there has
been no observational certainty yet, and the observational verification
of the theoretical expectation has been needed. In essence, just this
has been provided in this article using the exact statistical arguments.
All this means that the question  II. may be answered as follows: The
reason for this behavior {\it can be} instrumental. The instruments on
the Swift satellite are more sensitive than the other GRB detectors, and
the fainter GRBs should indeed be on average at higher redshifts.

Hence,
we allow to conjecture already now that the difference in the average
redshifts originates - either partly or fully - from instrumental
selection effects. 
For a more concrete answer a detailed study of the instrumental
properties of the considered satellites is needed, which is planned to
be provided in a forthcoming paper. It should quantify the importance of the 
instrumental biases.

Concerning the impact on the  redshift distributions, from the results
discussed above, three possibilities can be envisaged for the GRB
distribution:

A. The long subgroup of GRBs is unique (no further sub-grouping is
needed), and its redshift distribution is represented by the non-Swift
sample; the redshifts of the Swift sample are biased.  We believe that
this possibility is strongly disfavored, because the distribution of
GRBs before the Swift era were strongly biased (\cite{me04}), and in the
non-Swift sample only the brightest GRBs should be present. Hence, it is
hard to imagine that just the non-Swift sample represents the true
intrinsic redshift distribution of the unique long GRBs (\cite{me04}).
Also Berger et al. (2005) rejects this possibility.

B. The long subgroup of GRBs is unique, and its redshift distribution is
represented more or less well by the Swift sample; the redshifts of the
non-Swift sample are biased.  It is shown recently (\cite{ja05}) that
the Swift redshift data may represent the real intrinsic redshift
distribution of long GRBs, but one cannot exclude the possibility that
the exact GRB redshift distribution is neither properly represented by
the Swift sample nor by the non-Swift sample.  E.g. the redshift
distributions of the BATSE sample (\cite{MM96,scha01,lr02,bag03}) may be
quite different to that of the Swift sample.  Lin et al. (2004) even
claims that the majority of GRB in the BATSE sample are at $z>10$.
 
One should remark that there are 16 Swift GRBs with measured $z$ and
optical transient (OT) from the 32 Swift GRBs with OT observation
($50$\%), while for the non-Swift sample there are 38 GRBs with $z$ and
OT from the total of 65 GRBs with OT ($58$\%, \cite{gre05}).  (Seven
non-Swift events have measured $z$ without an OT.)  The remarkably
higher OT detection rate for the Swift bursts ($\simeq 43$\% vs. the
non-Swift $\simeq 29$\%) indicates that the lower redshift measurement
success ratio could be an indication of an observational bias from the
cosmic reddening: i.e.  for the significantly higher Swift redshifts a
spectral line may drop out of the V and R bands, and are detectable only
in the J, K bands or at even longer wavelengths. This clearly reduces
the success of the redshift determination.  The OT search efficiency  is
clearly $<100$\% due to observational constraints, so the $\simeq 43$\%
rate of the OT detection in the Swift era may suggest that the majority
of the GRBs are within the optically transparent cosmological region.

C. The long subgroup of GRBs is not necessarily unique, i.e. the different
redshifts in our samples may originate from different types of sources with
different redshift distributions: e.g. the non-uniqueness of the long subgroup
is also an alternative (\cite{bor04}).  Also Berger et al. (2005) considers, as
a possibility, that the Swift GRBs with higher redshifts represent a luminous
subgroup of the long class alone.  On the other hand, recent statistical
analyses show (\cite{ho05}) that three subgroups of GRBs are enough to explain
the statistical properties of the BATSE sample. Hence the long subclass should
not be further divided. It means that the observed differences might be
accounted only for the different observational strategies of the different
experiments.

However, simply the different energy sensitivity of the different spacecrafts
could not explain satisfactorily the differences in the redshifts.
The Swift and non-Swift trigger energy ranges overlap: in the non-Swift
samples' there are 27 ``softer'' and 6 ``harder'' GRBs (``softer'' and
``harder'' means that the observed trigger energy is below or above the average
Swift trigger energy), while 12 GRBs' trigger energy was within the Swift band.
E.g. only $\approx 13$\% of GRBs were triggered above the Swift energy band,
and the median $z$ of this small subsample is higher ($z_{\mbox{\em med}}=
1.64$) than the non-Swift median $z_{\mbox{\em non-Swift,med}}= 1.02$. The
higher trigger energy may select GRBs of higher redshifts, however, this effect
could be even a small sample effect only.

Thus, in our opinion, both the B and C cases are quite possible, and a further
detailed study is needed.  This we plan to provide in a separate paper.  Here
we can state only as an answer to question III. that our statistical arguments
relying on the observational data alone suggest that the average redshifts of
the long GRBs can be great indeed ($z \simeq (2.4-2.6)$ or even larger). 

\begin{acknowledgements}

This study was supported by Hungarian OTKA grant T48870 and NASA grant
NAG5-13286. Useful remarks of an anonymous referee are acknowledged.

\end{acknowledgements}

\end{document}